\documentclass{article}

\begin{document}

\centerline { \bf A POLYNOMIAL WEYL INVARIANT SPINNING MEMBRANE ACTION 
}

\bigskip

\centerline { Carlos Castro}

\bigskip

\centerline { Center for Theoretical Studies of Physical Systems}
\centerline { Clark Atlanta University, Atlanta, GA. 30314 }
\centerline{ November 2002 }
\bigskip

\centerline { \bf Abstract }

\bigskip
A review of the construction of a Weyl-invariant spinning-membrane
action that is $polynomial$ in the fields,  without a cosmological 
constant  
term,  comprised of quadratic and quartic-derivative terms,  and where 
supersymmetry is linearly realized, is presented.  The action is 
invariant 
under a $modified$ supersymmetry transformation law which is derived 
from a 
new $ Q + K + S $ sum-rule based on the  $3D$-superconformal algebra .

\bigskip

A satisfactory spinning membrane Lagrangian has not been
constructed yet, as far
as we know. Satisfactory in the sense that a suitable
action must be one which is $polynomial$ in the fields,
without (curvature) $R$ terms which interfere with the
algebraic  elimination of the three-metric, and also where
supersymmetry is linearly realized in the space of
physical fields.

Lindstrom and Rocek [1] were the first ones to
construct a Weyl invariant spinning membrane action.
However, such action was highly $non-polynomial$
complicating the quantization process . The suitable action to
supersymmetrize is the one of
Dolan and Tchrakian (DT) [2] without a cosmological constant
and with quadratic and quartic-derivative terms.
Such membrane action is basically a Skyrmion like action.

We shall write down the supersymmetrization of the polynomial
$DT$ action that is devoid of
$R$ and kinetic gravitino terms; where supersymmetry is
linearly realized and the gauge algebra closes [3].
The crux of the work [3] reviewed  here lies in the
necessity to Weyl-covariantize the Dolan--Tchrakian action
through the introduction of extra fields. These are the gauge
field of  dilations , $b_\mu$, and the scalar coupling , $A_0$ of
dimension $(length)^3$, that must appear in front of the quartic
derivative terms of the DT action.

Our coventions are: Greek indices stand for
three-dimensional ones; Latin indices for spacetime ones :
$i,j =0,1,2....D$. The signature of the $3D$ volume is
$(-,+,+)$. The Dolan--Tchrakian Action for the bosonic $p$-brane
( extendon) with vanishing cosmological constant
in the case that $p=odd; p+1=2n$ is :

$$ L_{2n} =\sqrt {-g} g^{\mu_1 \nu_1}.... g^{\mu_n \nu_n}
\partial_{[\mu_1} X^{i_1}.........\partial_{\mu_n]} X^{i_n}
\partial_{[\nu_1} X^{j_1}.........\partial_{\nu_n]} X^{j_n}
\eta_{i_1 j_1}.... \eta_{i_n j_n}.  . \eqno ( 1 ) $$

$\eta_{ij}$ is the spacetime metric and $g^{\mu\nu}$ is the
world volume metric of the $2n$ hypersurface spanned by the
motion of the $p-brane$. Antisymmetrization of indices is also
required. Upon the algebraic elimination of the world volume
metric $g_{\mu\nu}$ from the action and after pulgging its
value back into the action one recovers the Dirac--Nambu--Goto
action :

$$L_{2n} =\sqrt {-det( \partial_{\mu}
X^{i}(\sigma)\partial_{\nu} X^{j}(\sigma)\eta_{ij})} .  \eqno ( 2 ) $$

When $p=even$ , $p+1$ odd, a Lagrangian with zero
cosmological constant can also be constructed, however,
conformal invariance is $lost$.
In the membrane's case one has : $L=L_4 +aL_2.$; where $a$ is
a constant and $L_2$ is the standard
Howe-Polyakov-Tucker  quadratic actions
$ (\partial_{\mu} X^{i}(\sigma))^2$ and the quartic derivative
terms are the Skyrmion-like terms  :

$$ L_4 =\sqrt {-g} g^{\mu\nu} g^{\rho  \tau}\partial_{[\mu} 
X^{i}(\sigma)\partial_{\rho ]}
X^{k}(\sigma)\partial_{[\nu} X^{j}(\sigma)\partial_{\tau ]}
X^{l}(\sigma)  \eta_{ij} \eta_{kl}.  \eqno ( 3 ) $$

The Supersymmetrization of the Kinetic Terms is similar to the
construction of actions for the $3D$ Kinetic matter
superconformal multiplet where supersymmetry is linearly
realized and without $R$ terms.
In particular, we will show why the action is invariant under
a $modified$ ${\cal Q} $-transformation ,
${\tilde \delta}_Q = \delta_Q+ \delta'$, which includes a
compensating $S$ and $K$-transformation which cancel the
anomalous/spurious contributions to the ordinary $Q$ transformations ,
due to the explicit breakdown of $S$ and $K$-invariance of the
action,  which in turn, induce a breakdown of the ordinary $Q$
supersymmetry as well.

The final action is invariant under $ P, D, M^{ab}$
transformations : translations, dilations, Lorentz.
But is $not$ invariant under conformal boost $K$ and
$S$-supersymmetry.  The gauge algebra closes and the  action of the 
modified
supercovariant derivative operator  $ { \cal D }^c_\mu $  can be read 
from 
the commutator of two modified supersymmetry transformations :

$$[{\tilde \delta}_Q, {\tilde \delta} _Q ]  ( .  ) =
{\bar \epsilon^2} \gamma^\mu \epsilon^1 {\cal D}^c_\mu  ( . ) . \eqno ( 
4 ) 
$$
the  modified ${\cal D}^c_\mu$  supercovariant
derivative operator is obtained after a
compensating transformation $\delta'$ is added to the standard
$Q$-transformation to cancel the
anomalous $S$ and $K$-transformations of the supermultiplets
used to construct the action.

The scalar and kinetic multiplet of simple conformal SG in
$3D$ are respectively [4] :

$$ \Sigma_c =(A,\chi,F). ~~~
T_c(\Sigma_c)=(F,\gamma^\mu D_\mu^c \chi, \triangle A). \eqno ( 5 ) $$

For  a supermultiplet of Weyl weight $\omega (A)$, the supercovariant
derivatives and the D'Alambertian are respectively :

$$  D^c_\mu A =\partial_\mu A -{1\over 2} {\bar \psi}_\mu\chi  -\omega 
(A) 
b_\mu A.
\eqno ( 6 ) $$

$$  D^c_\mu \chi =(D_\mu -(\omega (A) +{1\over  2}  ) b_\mu )\chi -
{1\over 2} \gamma^\mu D_\mu^c A  \psi_\mu  - {1\over 2} F \psi_\mu -
\omega (A)  A \phi_\mu.  \eqno ( 7 ) $$

$$  \triangle A =D^c_a D^{ca} A=e^{-1}
\partial_\nu (eg^{\mu\nu}D^c_\mu A) +
{1\over 2} {\bar \phi}_\mu \gamma^\mu \chi -[\omega (A) - 1]
b^\mu D^c_\mu A +  $$
$$ 2\omega(A) A f^a_\mu e^\mu_a -{1\over 2} {\bar \psi}^\mu D^c_\mu 
\chi
-{1\over 2}{\bar \psi}^\mu \gamma^\nu \psi_\nu D^c_\mu A.  \eqno ( 8 )  
$$

The generalized spin connection
$\omega^{mn}_\mu ( e^m_\mu, b_\mu, \psi_\mu)$ is :

$$ \omega^{mn}_\mu =  - \omega^{mn}_\mu ( e ) - \kappa^{mn}_\mu ( \psi 
) +
e^n_\mu b^m  - e^m_\mu b^n. \eqno ( 9 )  $$

$$\kappa^{mn}_\mu = { 1 \over 4 } (  { \bar \psi_\mu } \gamma^m \psi^n
- { \bar \psi_\mu } \gamma^n \psi^m + { \bar \psi^m } \gamma_\mu 
\psi^n ).
\eqno ( 10 ) $$

The gravitino field strength is :

$$ \phi_\mu =
{ 1 \over 4 }  \sigma^{ \lambda \sigma} \gamma_\mu S_{ \sigma \lambda} 
.
\eqno ( 11 )  $$

$$ S_{\mu\nu}  = ( D_\nu + { 1 \over 2 } b_\nu ) \psi_\mu  - \mu 
\leftrightarrow \nu .
\eqno ( 12 ) $$

$$ e^{a \mu} f_{a \mu } = - { 1 \over 8 } R ( e, \omega ) -
{ 1 \over 4 } { \bar \psi_\mu} \sigma^{\mu\nu} \phi_\nu. \eqno ( 13 )  
$$
the variations of the gauge and matter
fields under $D$ ( dilatations ) ,  and $ Q, K, S, $ transfomations  is 
:

$$  \delta_D e^m_\mu =  - \lambda e^m_\mu. ~~~
\delta_D A = { 1\over 2 } \lambda A. ~~~ \delta_D F = { 3 \over 2 } 
\lambda 
F .
\eqno ( 14 )  $$
In general, a Weyl supermultiplet of weight $ \omega $  transforms 
under 
Weyl scalings  as :

$$ \delta_D A = \omega \lambda A. ~~~ \delta_D  \chi = ( \omega + { 1 
\over 
2 } ) \lambda \chi. ~~~ \delta_D  F = ( \omega + 1 ) \lambda F . \eqno 
( 15 
) $$
where $ \lambda$ is the Weyl scaling's infinitesimal parameter.
Under $ Q$-supersymmetry :

$$ \delta_Q A = { \bar \epsilon} \chi. ~~~ \delta_Q \chi = F \epsilon + 
\gamma^\mu D^c_\mu A \epsilon . ~~~ \delta_Q F = { \bar \epsilon} 
\gamma^\mu 
D^c_\mu \chi. \eqno ( 16 )  $$

$$ \delta_Q e^m_\mu = {\bar \epsilon } \gamma^m \psi_\mu. ~~~ \delta_Q 
\psi_\mu =
2 ( D_\mu + { 1 \over 2 } b_\mu ) \epsilon. ~~~ \delta_Q b_\mu = 
\phi_\mu .
\eqno ( 17 ) $$
Under $ S$-supersymmetry :

$$ \delta_S e^m_\mu = 0. ~~~ \delta_S \psi_\mu = - \gamma_\mu 
\epsilon_s.
~~~ \delta_S b_\mu = - { 1 \over 2 } \psi_\mu \epsilon_s . \eqno ( 18 ) 
$$

$$ \delta_S \omega^{mn}_\mu = - { \bar \epsilon_s } \psi_\mu. ~
\delta_S A = 0 . ~ \delta_S \chi = \omega ( A ) A \epsilon_s. ~~ 
\delta_S F 
=
(  1  - { 1 \over 2 } \omega ( A ) )  { \bar \chi} \epsilon_s . \eqno ( 
19 ) 
$$
The canonical scalar supermultiplet is $ inert$ under conformal boosts 
$ K$ 
and so are the $ e^m_\mu , \psi_\mu $. The $ b_\mu$ is $not$ inert :

$$ \delta_K  b_\mu = - 2 \xi_K^m e_{m \mu } . ~~~ \delta_K 
\omega^{mn}_\mu =
2 ( \xi_K^m  e^n_\mu  -  \xi_K^n e^m_\mu ) .  \eqno ( 20 ) $$

A superconformally invariant action for the kinetic terms in
$D=3$ requires to include $only$ the
following kinetic multiplet that is fully superconformaly
invariant :  $\Sigma_C\otimes T_C(\Sigma_C)$. The superconformal 
invariant
action was given in [4].

However, in this work we shall use the different combination  of 
supermultiplets

$$\Sigma^i_C \otimes T_C(\Sigma^j_C ) +  T_C(\Sigma^i_C)\otimes 
\Sigma^j_C 
-T_C( \Sigma^i_C \otimes  \Sigma^j_C). \eqno ( 21 ) $$
combination which happens to be the correct one to
dispense of the $R$ and kinetic gravitino terms. However this
combination $breaks$ explicitly the $S$ and
conformal boosts.  In doing so, it also  will break the
$Q$-invariance of any action  constructed based on such `` anomalous `` 
kinetic multiplet  [ 3 ].

The explicit components of the suitable combination of
multiplets which removes the $R$ and gravitino kinetic terms
and breaks $K$-symmetry and $S$-supersymmetry ,

$$\Sigma^i_C  \otimes T_C(\Sigma^j_C ) + T_C(\Sigma^i_C  )\otimes 
\Sigma^j_C 
-T_C( \Sigma^i_C \otimes \Sigma^j_C) . \eqno ( 22 ) $$
are  :

$$ A_{ij}= \bar{\chi_i} \chi_j.~~ \chi_{ij}  =F_i\chi_j +F_j\chi_i 
+A_i\gamma^\mu D^c_\mu (\omega ={1\over 2} ) \chi_j  +A_j\gamma^\mu 
D^c_\mu
(\omega ={1\over 2})\chi_i -  $$

$$  \gamma^\mu D^c_\mu (\omega =1)[A_i\chi_j +
A_j\chi_i].   $$

$$ F_{ij}= A_i \triangle (\omega ={1\over 2})
A_j +A_j \triangle (\omega ={1\over 2})A_i +2F_iF_j  -{\bar \chi}_i 
\gamma^\mu D^c_\mu ( \omega ={1\over 2})\chi_j  - $$

$$ {\bar \chi}_j \gamma^\mu D^c_\mu ( \omega
={1\over 2})\chi_i - \triangle ( \omega =1)[A_iA_j]  . \eqno ( 23 ) $$

The explicit expression for the multiplet components was given
in [3].  We can explicitly verify [3] that there is no gravitino  
kinetic 
term  and no curvature scalar terms, as expected. There  is an explicit 
presence of the $b_\mu$ terms,  which signals an  explicit breaking of 
conformal boosts.
Therefore, eliminating
the $R$ and gravitino kinetic terms is $not$ compatible with
$S$-supersymmetry nor $K$-symmetry.

Because the component $T_C(\Sigma_C\otimes\Sigma_C)$ does not
have the correct $S; K $ transformations laws,
the components of the latter supermultiplet $ (A_{ij};  ~\chi_{ij}; ~ 
F_{ij} 
) $
$do ~ not$ longer transform properly under ( ordinary ) $Q$
transformations because the supercovariant derivatives acting
on the $A_{ij}, \chi_{ij}$ acquire anomalous/spurious terms
due to the explicit $S$ and $K$-breaking pieces [3]. To correct
this problem one needs to modify the $Q$ supersymmetry by
adding  compensating $S,K$ transformations as shown below.

The supersymmetric quadratic derivative terms ( kinetic
action )  of the action are obtained by plugging in directly the 
components
$A_{ij},\chi_{ij}, F_{ij}$ into the following expression
while contracting the spacetime indices with $\eta_{ij}$ :

$$ {\cal L}_2=e\eta^{ij} [F_{ij} +{1\over  2}{\bar \psi}_\mu \gamma^\mu 
\chi_{ij} +
{1\over 2} A_{ij}{\bar \psi}_\mu \sigma^{\mu\nu} \psi_\nu ] . \eqno ( 
24 ) 
$$

After using a modified ${\cal Q} $-variation of the action the
spurious variations of the action, resulting from the $S$ and
$K$-symmetry breaking pieces in the definitions of
supercovariant derivatives,  will $cancel$ and the  action is invariant 
, up 
to total derivatives.
[3].  Using the modified derivatives, resulting from the
cancellation of the anomalos $S, K$ variations of the
the supermultiplet, the action  obtained from eq-( 24 ) will be 
invariant 
under the
following $modified$ ${\cal Q} $-transformations :

$$
{\tilde \delta} ^c_Q (A_{ij}) = \delta^c_Q (A_{ij}) = {\bar
\epsilon} \chi_{ij}. ~~  {\tilde \delta} ^c_Q (\chi_{ij}) = F_{ij} 
\epsilon 
+
\gamma^\mu  {\cal D}^c_\mu (A_{ij})\epsilon.  \eqno ( 25 ) $$

$$ {\tilde \delta} ^c_Q F_{ij} = {\bar \epsilon} \gamma^\mu {\cal  
D}^c_\mu 
(\chi_{ij}).
\eqno ( 26 )  $$
where now is essential to use the modified supercovariant
derivatives ${\cal D}^c_\mu$
associated with the modified ${\cal Q}$-transformations
laws in order to cancel the anomalous contributions stemming
from the $S$ and $K$-breaking pieces, since the
latter anomalous contributions to the $Q$-variations are
contained in the  $ordinary$ derivatives
$D^c_\mu (A_{ij}); ~ D^c_\mu  (\chi_{ij})$ .

Under ordinary $Q$-variations the Lagrangian density will
acquire spurious pieces [3] :
$(\Delta L)|_{spurious} = \delta_Q [ e L ]| _{spurious}$.
What the modified ${\tilde \delta_Q}{\cal S} =[ \delta_Q +  \delta' ] 
{\cal 
S}$ achieves is to have  $\delta' {\cal S} = - \Delta|_{spurious}$ so 
that 
the net $sum$  of
the anomalous $Q, K, S$ contributions cancels out giving zero at the 
end ;
i.e one has the correct $Q+K+S$ sum rule.

The modified covariant derivatives, appearing in the modified
${\cal Q}$ transformation laws of the supermultiplet of Weyl
weight $\omega = 2$, used to construct the action , have now
the appropriate form to ensure invariance of the actions under the 
modified
${ \tilde  Q} $ transformations  :

$${\cal D} ^c_\mu (A_{ij}) = \partial_\mu  A_{ij} - {1\over 2} {\bar 
\psi_\mu} \chi_{ij} -
2 b_\mu  A_{ij} .  \eqno ( 27 )  $$

$$ {\cal D} ^c_\mu \chi_{ij} = ( D_\mu - (2+  {1\over 2} ) b_\mu ) 
\chi_{ij} 
-
{1\over 2} \gamma^\nu {\cal D}^c_\nu (A_{ij}) \psi_\mu -
{1\over 2} F_{ij} \psi_\mu -  2  A_{ij} \phi_\mu .  \eqno ( 28 ) $$

Since ${\tilde \delta_Q } {\cal S} = 0$ ,  one has found the correct
$Q+K+S$ sum rule for the modified ${\cal Q}$ transformation.
Finally, we don't have $R$ terms, nor the kinetic terms for
the gravitino ; i.e there is no $f^m_\mu$ term.
${\cal Q} $-supersymmetry is linearly realized after the
elimination of $F^i$. No constraints arise after eliminating
the auxilary fileds $F^i$ [3].

The Supersymmetrization of the
Quartic Derivative Terms proceeds in analogous fashion .
Let us introduce the coupling-function supermultiplet,
$\Sigma_0=(A_0;\chi_0;F_0) $
whose Weyl weight is equal to $-3$ so that the tensor product
of $\Sigma_0$ with
the following multiplet, to be defined below, has a conformal
weight ,  $\omega=2$ as it is required in order to have Weyl invariant
actions. The following multiplet is the adequate one to build the
required supersymmetrization of the quartic derivative terms :

$$ K^{ijkl}_{\mu\nu\rho\tau} =K(\Sigma^i_\mu;\Sigma^j_\nu)
\otimes{T[K(\Sigma^k_\rho;\Sigma^l_\tau)}]  +({ij\leftrightarrow kl})~
and~({\mu\nu\leftrightarrow\rho\tau}) . \eqno ( 29 )  $$

$$T[K(\Sigma^i_\mu;\Sigma^j_\nu)\otimes  
{K(\Sigma^k_\rho;\Sigma^l_\tau)}].
\eqno ( 30 )  $$

$$K(\Sigma ,\Sigma)=\Sigma^i_C \otimes  T_C(\Sigma^j_C ) +
T_C(\Sigma^i_C  )\otimes \Sigma^j_C \equiv (A^K_{ij};~\chi^K_{ij}; 
~F^K_{ij}).
\eqno ( 31 ) $$

This combination of  these supermultiplets is the adequate one to
retrieve the DT action at the bosonic level and also ensures
that the $R$ terms do cancel from the final
answer. This is a similar case like the construction of the
quadratic terms which was devoid of $R$ and kinetic gravitino
terms. A similar calculation yields the components of the
supersymmetric-quartic-derivative
terms:

$$ A_{ijkl}= {\bar \chi}^K_{ij} \chi^K_{kl}.~~ ~ \chi_{ijkl} 
=F^K_{ij}\chi^K_{kl} +F^K_{kl}\chi^K_{ij} + $$

$$ A^K_{ij}\gamma^\mu D^c_\mu (\omega  ={2})\chi^K_{kl}  +
A^K_{kl}\gamma^\mu D^c_\mu (\omega ={2})\chi^K_{ij} - $$
$$  \gamma^\mu D^c_\mu (\omega =4)[A^K_{ij}\chi^K_{kl} +  
A^K_{kl}\chi^K_{ij}].  \eqno ( 32 ) $$

$$ F_{ijkl}= A^K_{ij} \triangle ( \omega = 2)
A^K_{kl} + A^K_{kl} \triangle (\omega = 2 ) A^K_{ij}
+2F^K_{ij} F^K_{kl} - $$

$$ {\bar \chi}^K_{ij} \gamma^\mu D^c_\mu ( \omega ={2})\chi^K_{kl} - 
{\bar 
\chi}^K_{kl}\gamma^\mu D^c_\mu  (  \omega = 2)\chi^K_{ij} -
\triangle ( \omega = 4)[A^K_{ij}A^K_{kl}].  \eqno ( 33 ) $$
where we have used the abreviations $A^K_{ij},\chi^K_{ij}$ and
$F^K_{ij}$ given by the well behaved kinetic multiplet.

Despite the fact that the defining quartic multiplet is devoid
of curvature terms we are once again faced with the anomalous
$Q$ transformations of the action due to the spurious
$S,K$-variations. To cure these anomalous variations of the
action we proceed exactly as before by using a modified ${\cal
Q}$ transformations that will cancel the spurious variations.

The complete ${\cal Q} $ supersymmetric extension of $L_4$
requires adding terms  which result as permutations of
${ijkl\leftrightarrow  ilkj\leftrightarrow  kjil\leftrightarrow klij}$ 
keeping
$\eta_{ij}\eta_{kl}$ fixed.  The action corresponding to the quartic 
derivative terms is :

$$   {\cal L}_4=  e\eta_{ij}\eta_{kl}~[A_0F^{ijkl}+F_0A^{ijkl} -{\bar  
\chi_0}\chi^{ijkl}+
{1\over 2}{\bar \psi}_\mu \gamma^\mu  (A_0\chi^{ijkl}+\chi_0A^{ijkl})+ 
$$

$$  {1\over 2}A_0A^{ijkl}{\bar \psi}_\mu  \sigma^{\mu\nu} \psi_\nu] .  
\eqno 
( 34 ) $$

The composite fields $A_{ijkl}, \chi_{ijkl}$ and $ F_{ijkl}$
are given in terms of the $A^K_{ij}, \chi^K_{ij}, F_{ij}$.
The action will be invariant under the $modified$ ${\cal 
Q}$-supersymmetry 
transformations laws,  which have the same structure as the variations 
of 
the quadratic derivatives multiplet :

$$  {\tilde \delta_Q} {\cal A}_{ijkl} = {\bar  \epsilon} 
\chi_{ijkl}.~~{\tilde \delta_Q}\chi_{ijkl}=  F_{ijkl} \epsilon + 
\gamma^\mu
{\cal D}^c_\mu {\cal A}_{ijkl}\epsilon.~~{\tilde \delta_Q}F_{ijkl}=
{\bar \epsilon} \gamma^\mu {\cal D}^c_\mu \chi_{ijkl}.  \eqno ( 35 )  
$$
where

$${\cal D} ^c_\mu (A_{ijkl}) = \partial_\mu A_{ijkl} -
{1\over 2} {\bar \psi_\mu} \chi_{ijkl} -  2 b_\mu A_{ijkl} . \eqno ( 36 
) $$
and similarly for :

$${\cal D} ^c_\mu \chi_{ijkl} = ( D_\mu - (2+ {1\over 2} )  b_\mu ) 
\chi_{ijkl}
- { 1 \over 2 } \gamma^\nu D^c_\nu ( A_{ijkl} ) \psi_\mu  -  { 1 \over 
2 } 
F_{ijkl} \psi_\mu
- 2 A_{ijkl}  \phi_\mu .    \eqno ( 36 ) $$.

Concluding,  we have presented  all the steps necessary to construct a 
satisfactory Weyl invariant spinning membrane action that is :
( i ) Polynomial in the fields  ( ii ) devoid of curvature terms ( iii 
) 
where supersymmetry is linearly realized  .  ( iv ) Upon setting the 
fermions to zero,  and eliminating the auxiliary fields, it yields the 
Weyl-covariant extension of the Dolan-Tchrakian action for the 
membrane.

\bigskip

\centerline{ \bf Acknowledgements}

We thank M. Bowers and J. Mahecha for their assistance.

\bigskip

\centerline{ \bf References }

[1] U.Lindstrom, M. Rocek, Phys. Letters {\bf B 218} (1988) 207;

[2] B.P.Dolan, D.H.Tchrakian, Physics Letters {\bf B 198} (1987) 447;

[3] C. Castro, `` Remarks on the existence of Spinning Membrane Actions ``
hep-th/0007031;

[4] T.Uematsu, Z.Physics {\bf C 29} (1985) 143-146 and {\bf C32} (1986)
33-42;

[5]. M. Duff, S. Deser, C. Isham , Nuc. Phys. {\bf B 114} (1976) 29;

[6] . E. Guendelman : `` Superextendons with a modified Measure ``
hep-th/0006079.

\end{document}